\documentclass[%
 reprint,
 superscriptaddress,
 amsmath,amssymb,
prl,
]{revtex4-2}

\usepackage{xr-hyper}
\usepackage{hyperref}

\usepackage{soul}
\usepackage{xcolor}
\usepackage{graphicx}
\usepackage{dcolumn}
\usepackage{bm}

\newcommand{\comment}[1]{}

\newcommand{\interior}[1]{%
  {\kern0pt#1}^{\mathrm{o}}%
}

\graphicspath{{./figure/}}

\begin{document}
\title{Strain Modulation Effects on the Topological Properties\\of a Chiral {\textit{\textbf p}}-Wave Superconductor}

\author{Yuto Shibata}%
\email{yuto.shibata@psi.ch}%
\affiliation{Condensed Matter Theory Group, Paul Scherrer Institut, 5232 Villigen PSI, Switzerland}%
\affiliation{Institute for Theoretical Physics, ETH Zürich, 8093 Zürich, Switzerland}
\author{Manfred Sigrist}%
\affiliation{Institute for Theoretical Physics, ETH Zürich, 8093 Zürich, Switzerland}

\date{\today}

\begin{abstract}
We present a study of strain modulation effects on electronic structures of a two-dimensional single-band chiral $p$-wave superconductor within the BCS mean-field scheme. We employ a lattice model and numerically solve the corresponding Bogolyubov-de Gennes equations. Assuming that strain modulation only modifies hopping amplitudes, we observe the emergence of local spontaneous supercurrents that are attributed to the non-trivial band topology of the system. We also report that strain modulation could induce the formation of topologically distinct domains within a single system, which is captured by spectral functions and local Chern markers.
\end{abstract}

\maketitle

\let\clearpage\relax
The application of strain to quantum materials is a clean and powerful way to probe their electronic properties. One of the advantages of this method is that strain can be continuously varied and, in case of uniaxial strain, its direction can be chosen arbitrarily in order to lower system symmetries. Recently, uniaxial strain has been applied to the unconventional superconductor Sr$_2$RuO$_4$ with the aim to obtain information on the superconducting pairing symmetry \cite{steppke2017strong}. In particular, the proposal that this superconductor may realize a topological superconducting state, such as the chiral $p$-wave phase, has been among the prime reasons for this investigation since the splitting of the superconducting transition into two consecutive ones would be expected. Although this separation into two transition has not been observed in thermodynamic measurements \cite{li2021heatcapacity}, $\mu$SR data detect a time reversal symmetry breaking transition well below the onset of superconductivity \cite{grinenko2021trsbstrain}. Recent NMR-Knight shift data are, however, in clear conflict with the spin-triplet chiral $p$-wave state \cite{pustogow2019NMR,ishida2020NMR} and initiated a renewed debate on the pairing symmetry in Sr$_2$RuO$_4$ \cite{kallin2012chiral,mackenzie2017even,mackenzie2020view}. In our study we will, nevertheless, use the chiral $p$-wave state as an example for a non-trivial topological superconducting order parameter with intriguing properties under the influence of symmetry-lowering strain. 

Recently, sample fabrication techniques based on focused ion beam techniques have been developed, which can be used to tailor inhomogeneous strain fields on a sub-micron scale \cite{bachmann2019spatial}. Combined with advanced probing techniques with spatial resolution such as scanning tunneling spectroscopy (STS) and angle-resolved photo-emission spectroscopy (ARPES) addressing electronic spectra, such specifically designed strain patterns can serve as diagnostic tools. In this paper, we study the behavior of a chiral $p$-wave superconductor under a regularly modulated uniaxial strain pattern, using a simple single-band lattice model which we treat by means of self-consistent Bogolyubov-de Gennes (BdG) theory. We will show how the spatial modulation and topology of this superconducting phase can lead to non-trivial structures in the electronic spectrum. 
\subsection{MODEL HAMILTONIAN}
In our study we aim at the description of a chiral $p$-wave superconductor subject to a spatial strain modulation. For this purpose it is convenient to model the superconductor by means of a single-band tight-binding formulation with nearest-neighbor pairing interaction on a square lattice, which is conveniently treated within the BdG formalism (see for example \cite{bouhon2014current}). 
We define the model Hamiltonian $\hat{\mathcal {H}}$ on a square lattice of $N_x \times N_y$ sites with periodic boundary conditions and denote the part of the tight-binding band structure by $\hat{\mathcal {H}}_0$ and the pairing interaction by $\hat{\mathcal {H}}_1$:
\begin{equation}
	\hat{\mathcal{H}} := \hat{\mathcal{H}}_0 + \hat{\mathcal{H}}_1. \label{eq:totalham}
\end{equation}
First, the pairing interaction $\hat{\mathcal{H}}_1$ can be derived from the coupling electron spin densities on nearest neighbor sites:
\begin{equation}
\hat{\mathcal{H}}_1 := -J \sum_{\langle i,j \rangle} \hat{\bm S}_i \cdot \hat{\bm S}_j, \label{eq:interacting}
\end{equation}
where $J >0 $ is the (ferromagnetic) coupling constant, and $i,j$ $\in \{ (x, y) \in \mathbb{N} \times \mathbb{N}\, | \, 0 \leq x < N_x, \, 0 \leq y < N_y \} =: \Lambda$ label the lattice sites. The spin density operator is defined as $\hat{\bm S}_i := \frac{1}{2}\sum_{s s'} \hat{c}^\dagger_{i,s} {\bm \sigma}_{s s'}\hat{c}_{i,s'}$, where $s,s' \in \{\uparrow, \downarrow\}$ are spin indices and $\bm \sigma$ are the Pauli matrices. Applying the completeness relation for the Pauli matrices, one can rewrite $\hat{\bm S}_i\cdot \hat{\bm S}_j$ as
\begin{multline}
\hat{\bm S}_i \cdot \hat{\bm S}_j = \frac{1}{4}\sum_{s_1, s_2, s_3, s_4} (2\delta_{s_1,s_4}\delta_{s_2,s_3} - \delta_{s_1, s_2}\delta_{s_3, s_4})\\
\hat{c}^\dagger_{i, s_1}\hat{c}_{i, s_2}\hat{c}^\dagger_{j, s_3}\hat{c}_{j, s_4}.
\label{eq:spin_part}
\end{multline}
The creation and annihilation operators of electrons at site $i$ with spin $s$ are denoted by $\hat{c}^\dagger_{i,s}$ and $\hat{c}_{i,s}$, respectively. Through Eq. (\ref{eq:spin_part}) we find that Eq. (\ref{eq:interacting}) gives rise to attractive interactions between neighboring electrons if they have the same spin. As we ignore the spin-orbit coupling and magnetic fields we can reduce our discussion to that of electrons of one spin only as the other is completely identical. Therefore, we omit the spin indices from now on and replace Eq. (\ref{eq:interacting}) by
\begin{equation}
\hat{\mathcal{H}}'_1 := -\frac{J}{4} \sum_{\langle i,j \rangle} \hat{c}^\dagger_i \hat{c}_i \hat{c}^\dagger_j \hat{c}_j \label{eq:interacting_pol}
\end{equation}
introducing spinless electron operators.
Using the standard mean-field decoupling scheme, we rewrite the interaction term $ \hat{\mathcal{H}}'_1 $ as
\begin{equation}
\hat{\mathcal{H}}'_{\Delta} := -\sum_{x,y}
\left(
\Delta^{\hat{x}}_{x,y} \hat{c}^\dagger_{x+1,y}\hat{c}^\dagger_{x,y} + 
\Delta^{\hat{y}}_{x,y} \hat{c}^\dagger_{x,y+1}\hat{c}^\dagger_{x,y}
\right) + \text{h.c.},
\end{equation}
where we have replaced $i$ by $(x,y)$, which will be more convenient in the later treatment, and introduced the mean fields $\Delta^{\hat{x}}_{x,y}$ and $\Delta^{\hat{y}}_{x,y} \in \mathbb{C}$ that correspond to the $p_x$ and the $p_y$ components of the order parameter, respectively:
\begin{align}
\Delta^{\hat{x}}_{x,y} &:= \frac{J}{4}\langle \hat{c}_{x,y} \, \hat{c}_{x+1,y} \rangle_0 \label{eq:gap1},\\
\Delta^{\hat{y}}_{x,y} &:= \frac{J}{4}\langle \hat{c}_{x,y} \, \hat{c}_{x,y+1} \rangle_0 \label{eq:gap2}.
\end{align}
Note that $\langle F(\hat{c},\hat{c}^\dagger) \rangle_0 \in \mathbb{C}$ is the ground state expectation value of the electron operator combination $F(\hat{c},\hat{c}^\dagger)$ because we restrict ourselves to the case of zero temperature. \\
Next, we turn to $\hat{\mathcal{H}}_0$ which consists of the nearest- and next-nearest neighbor hopping of the spinless electrons:
\begin{multline}
	\hat{\mathcal{H}}_0 :=
	-\sum_{x,y} \left\{
	  t^{\hat{x}}_{x,y}  \hat{c}^\dagger_{x+1,y} \hat{c}_{x,y}
	  +t^{\hat{y}}_{x,y}  \hat{c}^\dagger_{x,y+1} \hat{c}_{x,y} \right.\\
	  \left. + t^{\hat{d}}\left(
	  \hat{c}^\dagger_{x+1,y+1} \hat{c}_{x,y} + \hat{c}^\dagger_{x-1,y+1} \hat{c}_{x,y}
	  \right)
	  \right\} + \text{h.c.} \\
	  - \mu \sum_{x,y} \hat{c}^\dagger_{x,y} \hat{c}_{x,y} \,, \label{Model:Eq:ham_hopping}
\end{multline}
where the hopping amplitudes in the $x$-, $y$-, and diagonal directions are denoted by $t^{\hat{x}}_{x,y}$, $t^{\hat{y}}_{x,y}$, and $t^{\hat{d}} \in \mathbb{R}_{\geq 0}$, respectively, and in the last term we include the chemical potential $\mu$. \\
At this point we introduce the strain through the nearest-neighbor hopping matrix elements $t^{\hat{x}}_{x,y}$ and $t^{\hat{y}}_{x,y}$
\begin{equation}
	t^{\hat{x}}_{x,y} := 1.0 + \delta t_{x,y}, \;\; t^{\hat{y}}_{x,y} := 1.0 - \delta t_{x,y},
\end{equation}
where $\delta t_{x,y} \in \mathbb{R}$, whereas $t^{\hat{d}}$ remains unchanged. Note that we take the nearest-neighbor hopping matrix element in the absence of deformation as the energy unit. This anisotropy $\delta t_{x,y}$ accounts for changes in the overlap of the electronic orbitals when the crystal lattice is deformed uniaxially along the main axes. In this form it is straightforward to include spatial modulation of the strain through the position dependence of $\delta t_{x,y} $. \\
In the following we numerically diagonalize the mean field Hamiltonian, $ \hat{\mathcal{H}}_{\text{MF}} := \hat{\mathcal{H}}_0 +\hat{\mathcal{H}}'_{\Delta}$ and adapt the mean fields self-consistently requiring that Eq. (\ref{eq:gap1}) and (\ref{eq:gap2}) be satisfied at all lattice sites simultaneously. Furthermore, we impose Neumann periodic boundary conditions in both the $x$- and $y$-direction, such that no surfaces or interfaces are implemented in our system. For the homogeneous situation, including uniform uniaxial strain, the tight-binding model in the presence of Eq. \eqref{Model:Eq:ham_hopping} yields the following band structure in normal states:
\begin{equation}
\xi_{\bm k} = -2t^{\hat{x}}\cos{k_x} -2t^{\hat{y}}\cos{k_y} -4t^{\hat{d}}\cos{k_x}\cos{k_y} -\mu,
\end{equation}
with wave number $\bm k = (k_x, k_y)$ in the 1st Brillouin zone (BZ). In Fig. \ref{fig:Lifshitz} we show the Fermi surface for different values of the uniaxial deformation parameter $ \delta t$. We observe that there is a critical value $ \delta t^* $ where the Fermi surfaces changes topology passing through a Lifshitz transition leading to an open Fermi surface. 
\subsection{TOPOLOGY OF SUPERCONDUCTING PHASE}
The most stable superconducting phase resulting from the pairing interaction defined above is the chiral $p$-wave state where both bond mean fields, $\Delta^{\hat{x}}$ and $\Delta^{\hat{y}}$, are non-vanishing with a relative phase $\pm \pi /2$. This time reversal symmetry breaking state within a BdG Hamiltonian is a topologically non-trivial phase belonging to the symmetry class D in two-dimensions and is thus classified by the Chern number $\mathcal C \in \mathbb{Z}$ according to the periodic table of topological insulators and superconductors \cite{schnyder2008classification, ryu2010topological}. In uniform systems with $\mu \, (>0)$, $\delta t_{x,y} = \delta t \neq 0 \; \text{for all } (x,y) \in \Lambda$ and $t^{\hat{d}}=0$, the Chern number is given by \cite{sato2009topological, asahi2012topological}
\begin{align}
	\mathcal{C} = 
	\begin{cases}
		 1, & \;\text{if} \;\; |\delta t| < \mu/4, \\
		 0, & \;\text{if} \;\; |\delta t| > \mu/4. \\
	\end{cases}
\end{align}
The topological phase transitions at $\delta t = \pm \mu/4$ coincide with the Lifshitz transition of the Fermi surface which touches the BZ boundary at the points $(k_x, k_y) = (0,\pm \pi)$ or $(\pm \pi, 0)$. The transition is connected with a closing of the quasi-particle gap at BZ boundary. In $k$-space, the quasi-particle spectral gap has the form $ \Delta_{\bm k} = \Delta^{\hat{x}} \sin k_x + \Delta^{\hat{y}} \sin k_y $ and corresponds to $ \Delta_{\bm k} = \Delta_0 (\sin k_x \pm i \sin k_y) $ for the isotropic case, i.e. $\delta t = 0 $. Obviously, a simple change of the Fermi surface topology is sufficient to change the topological properties of the superconducting state without anomalous modification of the pairing state. For the open Fermi surface the state becomes topologically trivial (i.e. $ \mathcal{C} = 0$).
 \begin{figure}[tpp]
	\hspace*{-0.6cm}
	\includegraphics[width=75mm]{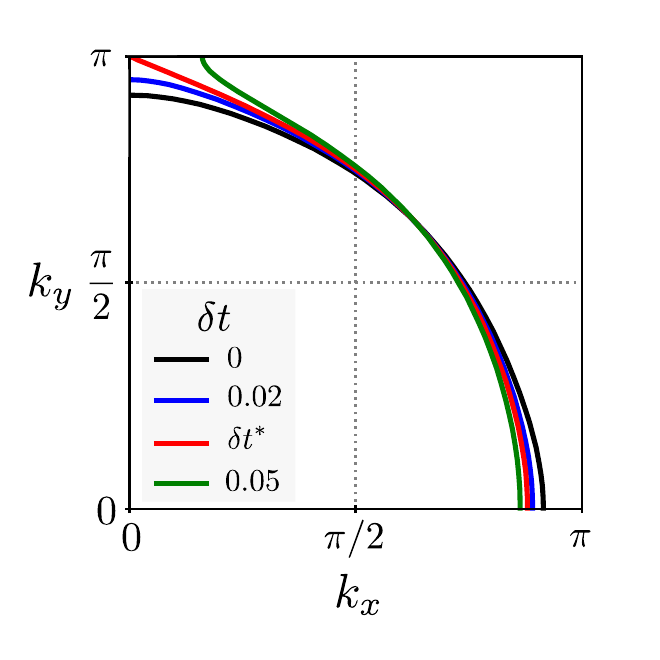}
	\vspace*{-0.2cm}
	\caption{Modification of the Fermi surface due to uni-axial strain of different strength $\delta t$ for models with $t^{\hat{d}} = 0.375$ and $\mu = 1.375$. An Lifshitz transition takes place at $(k_x, k_y) = (0, \pm \pi)$ when $\delta t = \delta t^* (= 0.03125)$.}
	\label{fig:Lifshitz}
\end{figure}
\\
The analogous behavior occurs also with non-vanishing $t^{\hat{d}}$ whose effect is to merely shifts the transition point to ${\delta t}^* = |\mu/4 - t^{\hat{d}}|$, corresponding to the Lifshitz point as seen in Fig. \ref{fig:Lifshitz}, 
\begin{align}
\mathcal{C} = 
\begin{cases}
1, & \;\text{if} \;\; |\delta t| < {\delta t}^*, \\
0, & \;\text{if} \;\; |\delta t| > {\delta t}^*. \\
\end{cases}
\end{align}
\subsection{SIMULATIONS AND DISCUSSIONS}
\begin{figure}[htbp]
	\hspace*{-0.6cm}
	\includegraphics[width=95mm]{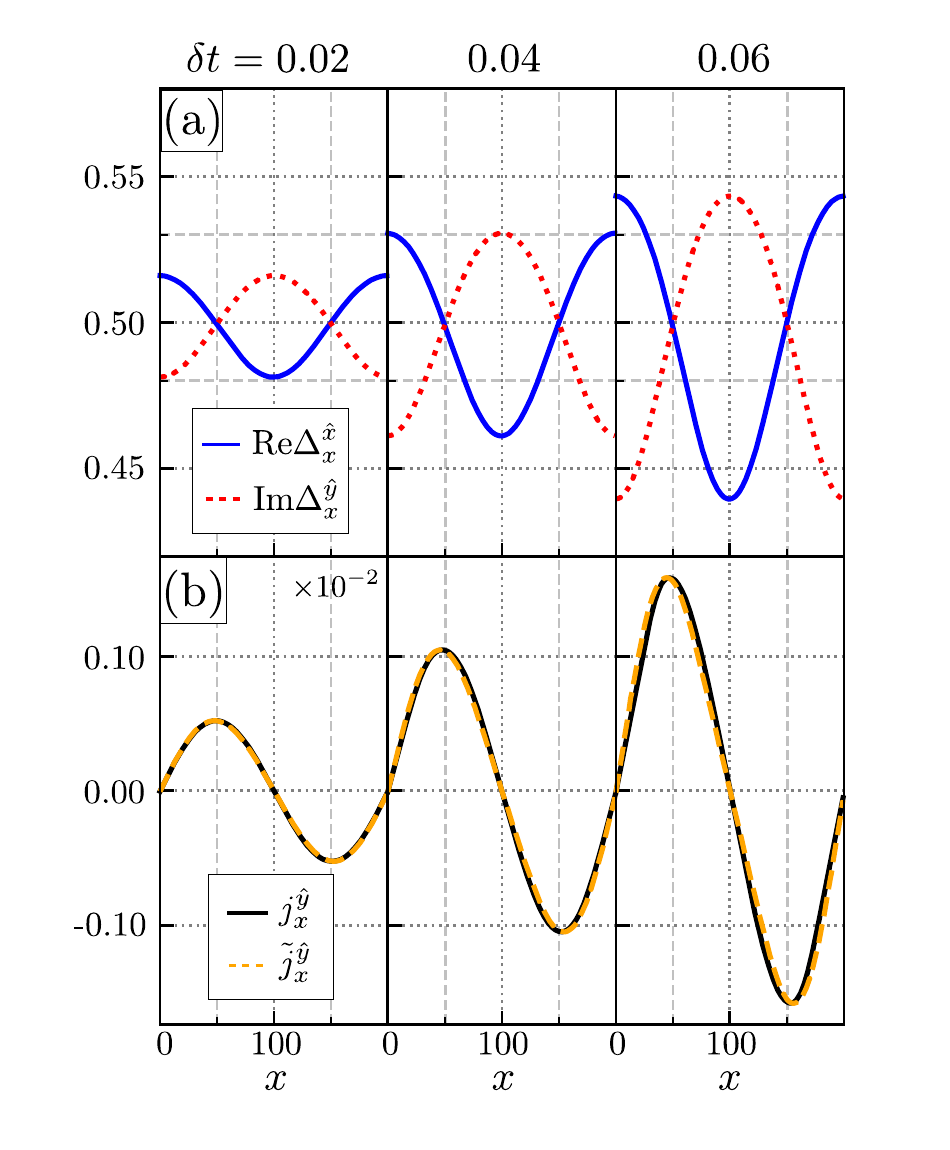}
	\vspace*{-1.2cm}
	\caption{Spatial dependence of (a) $\Delta^{\hat{x}}_x$ (in solid blue curves) and $\Delta^{\hat{y}}_x$ (in broken red curves) and (b) spontaneous charge currents $j^{\hat{y}}_x$ (in solid black curves) and the fitting curves $\tilde{j}^{\hat{y}}_x$ based on Eq. (\ref{eq:spontaneous_current_single_gamma}) (in broken orange curves) in the presence of strain modulation with different strain modulation amplitudes $\delta t = 0.02, 0.04$, and $0.06$. Note that $\text{Im} \Delta^{\hat x}_x$, $\text{Re} \Delta^{\hat y}_x$, and $j^{\hat x}_{x}$ are found vanishing for all the cases above. In addition, we set $e=\hbar=1$ in (b).}
	\label{fig:gap_current}
\end{figure}

\begin{figure*}[htbp]
	\vspace*{-0.8cm}
	\hspace*{-0.5cm}
	\includegraphics[width=190mm]{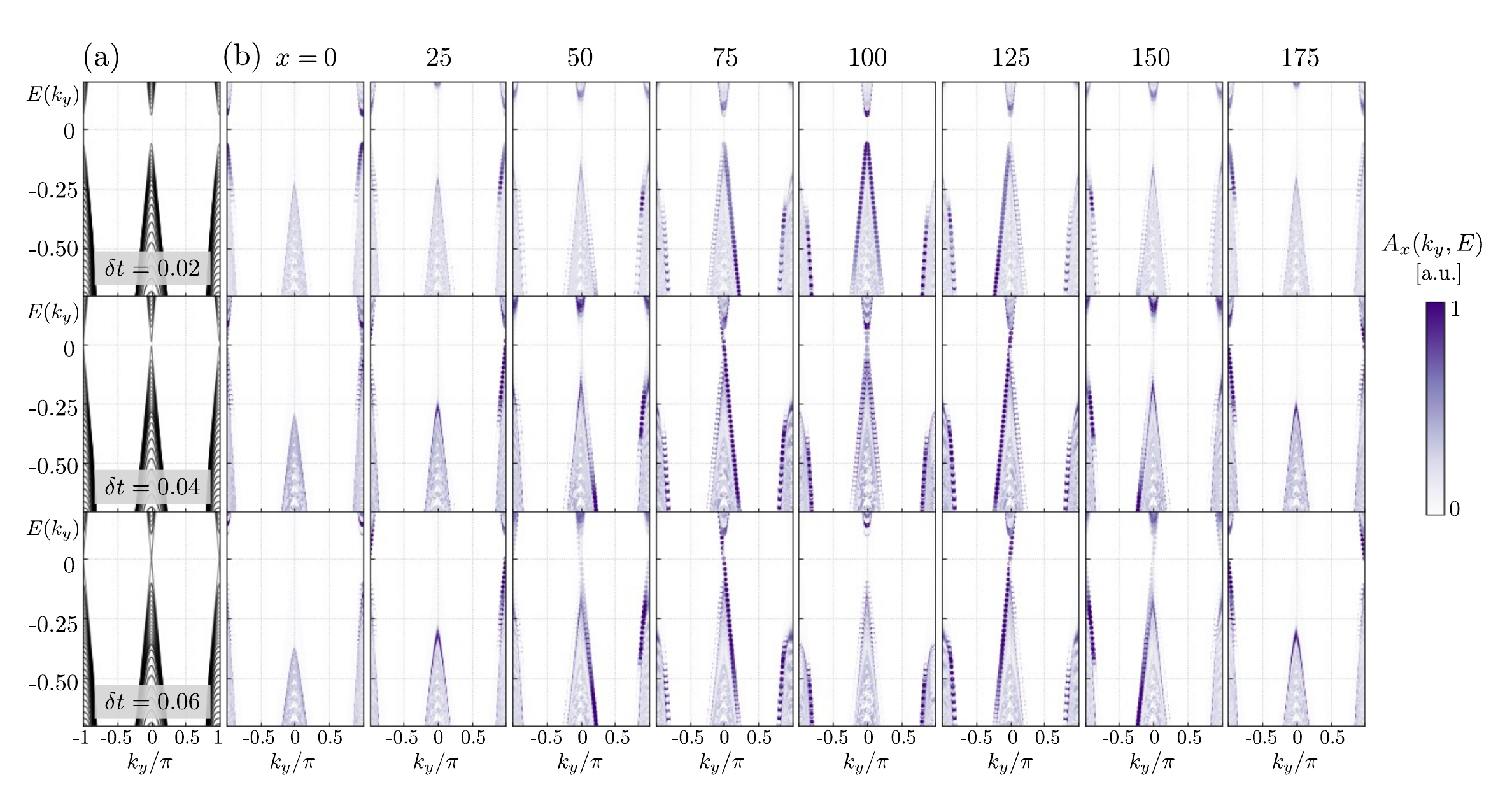}
	\vspace*{-0.5cm}
	\caption{Plots of (a) quasi-particle spectra and (b) spectral functions defined in Eq. (\ref{eq:spectral_function}) for systems with $\delta t = 0.02$ (top), $0.04$ (middle), and $0.06$ (bottom). The color bar is for (b), and $A_x(k_y, E)$ is normalized individually for each $x$. Gapless states are absent in the system with $\delta t = 0.02 < \delta t^*$, whereas four localized gapless states, or two pairs of chiral edge modes, are present in the systems with $\delta t = 0.04$ and $0.06 > \delta t^*$ around $(x,k_y) = (25,\pm\pi), (75,0), (125,0)$, and $(175,\pm\pi)$, respectively.}
	\label{fig:spectral_function}
\end{figure*}

\begin{figure*}[bhtp]
	\hspace*{-0.5cm}
	\includegraphics[width=180mm]{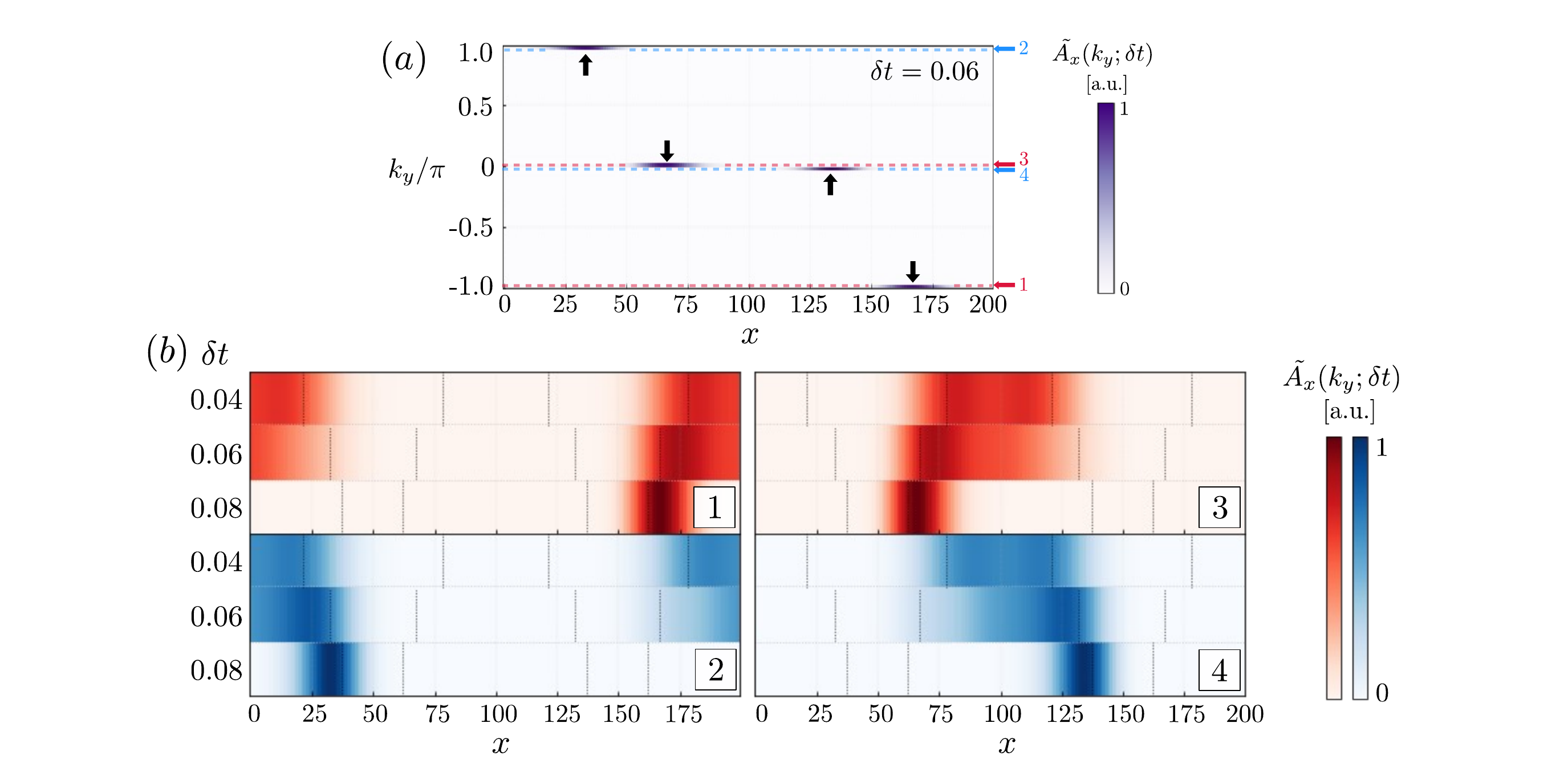}
	\vspace*{-0.2cm}
	\caption{(a) Integrated spectral function $\tilde{A_x}(k_y)$ defined in Eq. (\ref{eq:integrated_spectral_fucntion}) for $\delta t = 0.08$. The dotted horizontal lines are to indicate and label specific momenta $k_y = -(\pi-\epsilon), \pi - \epsilon, 0_+$, and $0_-$ by 1, 2, 3, and 4, respectively. For each of these momenta, the $\delta t$ dependence of $\tilde{A_x}(k_y)$ is shown in (b). The color of red (blue) is used if the group velocity of each of the gapless states, determined from the spectra in Fig. \ref{fig:spectral_function}, is negative (positive). The dotted lines in (b) indicate where $\delta t_x = \delta t^*$ for each $\delta t$ case.}
	\label{fig:boundary_positions}
\end{figure*}

\begin{figure}[tp]
	\hspace*{-0.55cm}
	\includegraphics[width=100mm,height=100mm]{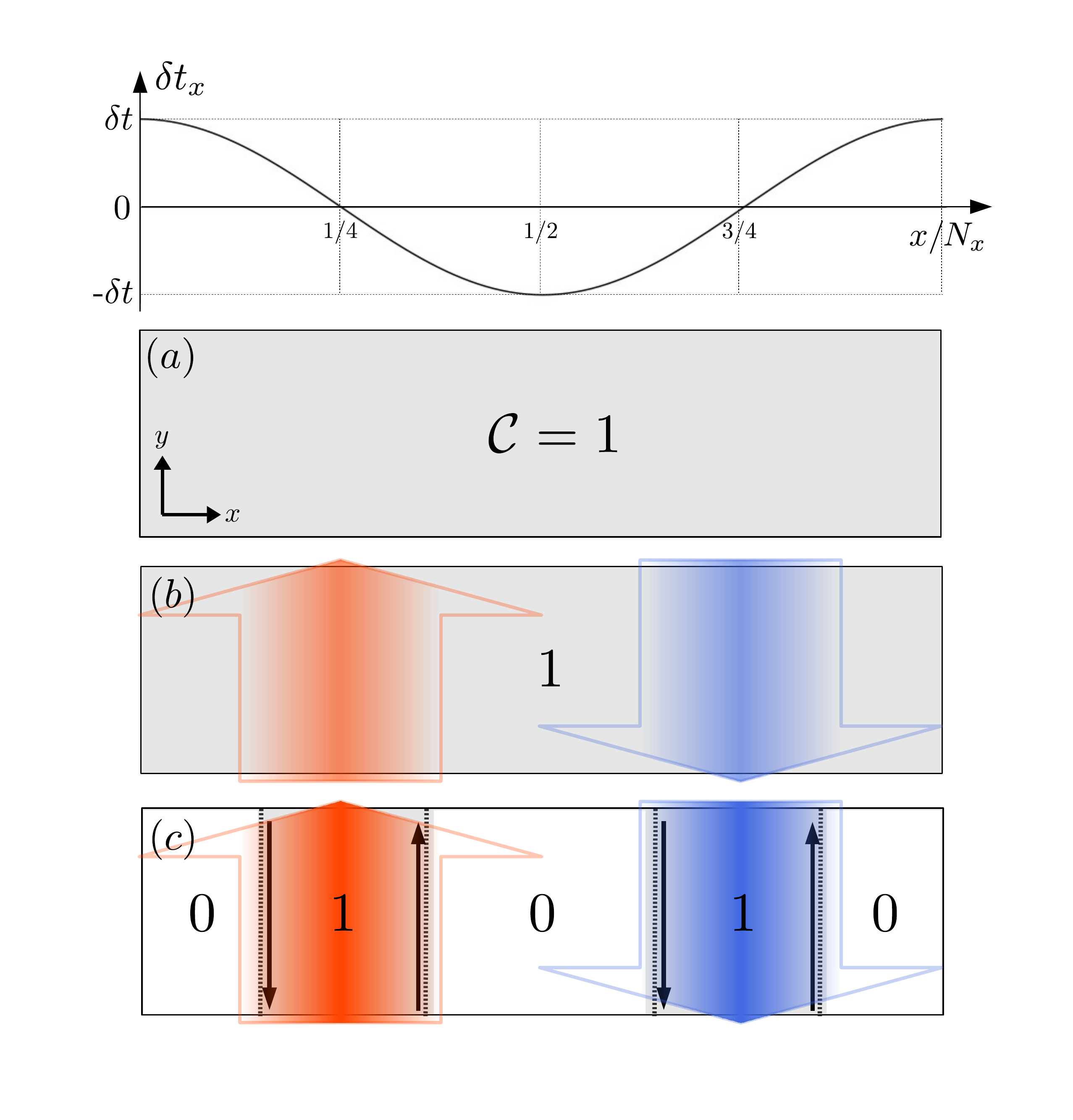}
	\vspace*{-0.9cm}
	\caption{(Top) Plot of spatial modulation $\delta t_x$ as a function of $x/N_x$. (a)-(c) Sketches of evolution of spontaneous charge currents, presented with arrows with their gradation indicating their magnitude and their color indicating their signs (blue for negative values). Also, illustrated are the formation of topologically distinct domains, whose Chern numbers are indicated by integers, as $\delta t$ increases; (a) $\delta t = 0$, (b) $0 < \delta t < \delta t^*$, and (c) $\delta t^* < \delta t$. Note that the presence of chiral edge states in (c) is deduced from the spectral functions Fig. \ref{fig:spectral_function} and is indicated by black arrows. Provided that the gapless modes have negative effective charge, they generate supercurrents in the direction to which the arrows point.}
	\label{fig:topo_domain}
\end{figure}

Now we consider the situation with a periodically modulated uniaxial strain and analyze its impact on the properties of the chiral $p$-wave superconductor by means of a full numerical evaluation of the BdG equation. For this purpose we use the system parameters,
\begin{align}
	\mu &= 1.375,\\
	t^{\hat{d}} &= 0.375,
\end{align}
which yield a closed nearly circular Fermi surface in the undistorted case (analogous to the $ \gamma $-band of Sr$_2$RuO$_4$). Thus, the Lifshitz transition occurs when ${\delta t} = 0.03125 =: {\delta t}^*$ (Fig. \ref{fig:Lifshitz}) in uniform systems. The spatial variation of the uniaxial strain is parametrized as
\begin{equation}
\delta t_{x,y} = \delta t\, \cos\left(2\pi x/N_x\right) \label{eq:strain_modulation_1},
\end{equation}
with a modulation amplitude $\delta t \geq 0$. For this periodic modulation along the $x$-axis we use a wavelength commensurate with $ N_x$ such that we still can use the Neumann periodic boundary conditions along both the $x$- and $y$-direction. Our self-consistent computational simulation of the superconducting phase results in the following three main observations which we will detail further below:
\begin{enumerate}
\item The modulation introduced in Eq. \eqref{eq:strain_modulation_1} with non-zero finite $\delta t$ leads to a spatial out-of-phase modulation of $\Delta^{\hat{x}}_{x,y}$ and $\Delta^{\hat{y}}_{x,y}$ along the $x$-direction (Fig. \ref{fig:gap_current}-(a)). 

\item Spontaneous supercurrents flow in the $y$-direction, which is perpendicular to the spatial modulation of the order parameters (Fig. \ref{fig:gap_current}-(b)).

\item  For $\delta t > {\delta t}^*$, gapless chiral modes dispersing along $ y$-direction appear locally, while the order parameters sustain non-zero finite values and the relative phase difference of $\pm \pi/2$ everywhere. This manifests the formation of domains \emph{with different Chern numbers} due to the deformation (Fig. \ref{fig:spectral_function}, \ref{fig:boundary_positions}, and \ref{fig:topo_domain}).
\end{enumerate}

In the following, the site index $y$ is omitted since we have imposed translational invariance along the $y$-direction. For the calculation we use a system size $N_x = 200$, $N_y = 500$, and the coupling strength $J=12.8$ such that the order parameters are non-vanishing in the whole parameter range explored below. We neglect any modulation of $ J$ due to strain, which would not change the results qualitatively.\\
\paragraph{Spatial Modulation of the Order Parameters} -- The strain modulation Eq. (\ref{eq:strain_modulation_1}) reduces the tetragonal crystal symmetry through its uniaxial nature and the loss of translational invariance along the $x$-direction. This leads to the lifting of the degeneracy of the two order parameter components $\Delta^{\hat{x}}_{x}$ and $\Delta^{\hat{y}}_{x}$. Their resulting spatial modulation along the $x$-axis is shown in Fig. \ref{fig:gap_current}-(a) for the three cases $\delta t = 0.02, 0.04,$ and $0.06$. The relatively large coupling $ J $ ensures a short coherence length (tightly bound Cooper pairs) and numerical accuracy. As a result the two components follow closely the modulation of strain fields. Both of the gap components remain non-zero finite and do not show any singularity. Note that $\text{Im}\Delta^{\hat{x}}_{x} = \text{Re}\Delta^{\hat{y}}_{x}$ = 0 at every site $x$ for the full range of $\delta t$ used here. In other words, the chiral $p$-wave states retain its phase structure even in the presence of the strain modulation.\\

\paragraph{Spontaneous Charge Currents} -- The out-of-phase spatial modulation of the two order parameter components leads to a spontaneous supercurrent pattern. We introduce the charge current density operator through the general relation,  
\begin{equation}
\hat{\bm j} := \frac{ie}{\hbar} \sum_{(x,y) \in \Lambda}\left\{ (x \bm e^{\hat{x}} + y \bm e^{\hat{y}} ) \,[\hat{c}^\dagger_{x,y} \hat{c}_{x,y}, \hat{\mathcal H}_0] \right \}, \label{eq:spontaneous_current_all}
\end{equation}
where $e$ is the elementary charge, and $\bm e^{\hat{x}}$ and $\bm e^{\hat{y}}$ are primitive vectors of the square lattice. Within our model, the two in-plane components of the current density operator are expressed as
\begin{multline}
 \hat{j}^{\hat{x}}_{x,y} :=
 -\frac{ie}{\hbar}
 \left\{
 t^{\hat{x}}_{x,y}
 \left(
 \hat{c}^\dagger_{x+1,y}\hat{c}_{x,y} - \hat{c}^\dagger_{x,y}\hat{c}_{x+1,y}
 \right) \right.\\
 \left.
 +t^{\hat{d}}
 \left(
  \hat{c}^\dagger_{x+1, y+1} \hat{c}_{x,y}
 +\hat{c}^\dagger_{x+1, y-1} \hat{c}_{x,y} \right.\right.\\
 \left.\left.
 -\hat{c}^\dagger_{x+1, y-1} \hat{c}_{x+1,y}
 -\hat{c}^\dagger_{x+1, y+1} \hat{c}_{x+1,y}
 \right)
 \right\},
 \label{eq:spontaneous_current_x}
\end{multline}
and
\begin{multline}
 \hat{j}^{\hat{y}}_{x,y} :=
 -\frac{ie}{\hbar}
 \left\{
 t^{\hat{y}}_{x,y}
 \left(
 \hat{c}^\dagger_{x,y+1}\hat{c}_{x,y} - \hat{c}^\dagger_{x,y}\hat{c}_{x,y+1}
 \right),\right.\\
 \left.
 +t^{\hat{d}}
 \left(
  \hat{c}^\dagger_{x+1, y+1} \hat{c}_{x,y}
 +\hat{c}^\dagger_{x-1, y+1} \hat{c}_{x,y} \right.\right.\\
 \left.\left.
 -\hat{c}^\dagger_{x+1, y} \hat{c}_{x,y+1}
 -\hat{c}^\dagger_{x-1, y} \hat{c}_{x,y+1}
 \right)
 \right\},
 \label{eq:spontaneous_current_y}
\end{multline}
such that 
\begin{equation}
\hat{\bm j} = \sum_{x,y} \left( \hat{j}^{\hat{x}}_{x,y} \bm e^{\hat{x}} + \hat{j}^{\hat{y}}_{x,y} \bm e^{\hat{y}} \right).
\label{eq:spontaneous_current_decomposition}
\end{equation}
From Eq. \eqref{eq:spontaneous_current_x}-\eqref{eq:spontaneous_current_decomposition}, it can be shown that Kirchhoff's current law is satisfied at every site.\\
For the spatially modulated superconducting phase observed in Fig. \ref{fig:gap_current}-(a), $j^{\hat{x}}_x := \langle \hat{j}^{\hat{x}}_x \rangle_0$ is found to vanish. On the other hand, $j^{\hat{y}}_x := \langle \hat{j}^{\hat{y}}_x \rangle_0$ shows an alternating current density in accordance with the order parameters as displayed in Fig. \ref{fig:gap_current}-(b). In particular, $j^{\hat{y}}_x$ vanishes at the extrema of the order parameters and are maximal where the order parameters vary most strongly in space. The spatial profile of $j^{\hat{y}}_x$ fits well with the supercurrent predicted from a Ginzburg-Landau formulation where the relevant expression \cite{sigrist1991phenomenological,huang2015nontopological}:
\begin{equation}
	\tilde{j}^{\hat{y}}_{x} := \gamma(\bar{\Delta}^{\hat y}_x \partial_x \bar{\Delta}^{\hat{x}}_{x} - \bar{\Delta}^{\hat x}_x\partial_x \bar{\Delta}^{\hat{y}}_{x}), \label{eq:spontaneous_current_single_gamma}
\end{equation}
where $\bar{\Delta}^{\hat{x}}_{x} := \text{Re}{\Delta}^{\hat{x}}_{x}$, $\bar{\Delta}^{\hat{y}}_{x} := \text{Im}{\Delta}^{\hat{y}}_{x}$, and the coefficient $\gamma$ is a site-independent phenomenological parameter which in principle can be derived from the microscopic model. In Fig \ref{fig:gap_current}-(b) we use
$ \gamma $ as the only fitting parameter and observe a surprisingly good match with the numerical data, using the order parameter displayed in Fig \ref{fig:gap_current}-(a). A slight deviation appears with increasing $ \delta t$, which suggests the presence of further contributions that are not resolved within the Ginzburg-Landau expression.\\
It is interesting to consider this feature in the context of the attempts to observe edge currents in Sr$_2$RuO$_4$, which have failed so far \cite{kirtley2007upper,curran2014search}. 
This could be attributed to various effects, in particular, band structure effects and surface scattering conditions for the electrons which suppress the current flow \cite{bouhon2014current,huang2015nontopological, etter2018spontaneous}. The latter is not an issue for the currents possible in our case, such that the conditions to find spontaneous currents may be better in this case of modulated uniaxial strain.
\paragraph{Topological Domains} --
Our simulation suggests that, when $\delta t > {\delta t}^*$, our system should incorporate topologically distinct domains through the change of Fermi surface topology. In the following we substantiate this by two approaches. The first considers the local spectral function of electronic states, and the second is connected with the concept of \emph{local Chern markers}.\\
First, we here define the local spectral function $A_x$ at site $x$ as a function of the momentum $k_y$ and the energy $E$ as
\begin{equation}
A_x(k_y,E) := -\frac{1}{\pi}\text{Im}G^R_{x,x}(k_y,E),
\label{eq:spectral_function}
\end{equation}
with the local retarded Green's function $G^R_{x,x'}(k_y, E)$,
\begin{multline}
G^R_{x,x'}(k_y, E) := \langle \hat{c}^\dagger_x(k_y) \frac{1}{E - \hat{\mathcal{H}}_\text{MF} + i \eta} \hat{c}_{x'}(k_y)\rangle_0 \\
+ \langle \hat{c}_{x'}(k_y) \frac{1}{E + \hat{\mathcal{H}}_\text{MF} + i \eta} \hat{c}^\dagger_x(k_y)\rangle_0,
\end{multline}
where $ \eta \to 0_+$. Note that we only use the position label $x$ and suppress $y$ in our notation as the spectrum is translation invariant along the $y$-direction. Fig. \ref{fig:spectral_function} demonstrates the $ k_y$-dependence of the complete quasi-particle spectrum and of the local spectral function at several different sites $x$ for systems with $\delta t = 0.02$ (top), $0.04$ (middle), and $0.06$ (bottom). 
We find that the quasi-particle spectrum remains fully gapped everywhere for $\delta t = 0.02 < \delta t^*$, whereas gapless states emerge at $k_y=0,\pm\pi$ for $\delta t = 0.04$ and $0.06 > {\delta t}^*$. In order to examine where those gapless states appear in real space and their $\delta t$ dependence, we define and evaluate integrated spectral function $\tilde{A_x}(k_y)$ over a small energy interval $E \in [-0.1,0]$ as
\begin{equation}
	\tilde{A}_x(k_y) := \int^{0}_{-0.1} \text{d}E \, A_x(k_y, E),\label{eq:integrated_spectral_fucntion}
\end{equation}
which captures the contribution of the gapless states to $A_x(k_y,E)$, thus enabling us to identify the spatial locatations of the gapless modes. Shown in Fig. \ref{fig:boundary_positions}-(a) is $\tilde{A}_x(k_y)$ for $\delta t = 0.06$. We observe the spatially localized four gapless modes each of which is associated with a specific momentum $k_y$. Fig. \ref{fig:boundary_positions}-(b) shows the $\delta t$ dependence of $\tilde{A_x}(k_y)$ at $k_y = \pm(\pi - \epsilon)$ and $0 \pm \epsilon$ with $\epsilon \rightarrow 0_+$. The gapless modes are observed to emerge and localize around where $x$ satisfies $\delta t_x = \delta t^*$, as indicated in \ref{fig:boundary_positions}-(b) by vertical dotted black lines for each $\delta t$. As seen in Fig. \ref{fig:spectral_function}-(b), these local modes are chiral, that is each has a single zero-energy-crossing with a specific velocity in the $y$-direction and localized at the boundary of two superconducting domains that are topologically distinct from each other. 
For example, for $\delta t = 0.06$, we expect two topologically non-trivial domains with $\mathcal{C} = 1$ appearing between $x=32$ and $67$ as well as between $x=137$ and $167$. In the whole range, $\delta t_x$ changes continuously and satisfies the condition for a closed Fermi surface, $|\delta t_x| < \delta t^*$, in these two intervals each. Then we expect two topologically trivial domains ($\mathcal{C} = 0$) to be present in the in-between region, where $|\delta t_x| > \delta t^*$ holds (see Fig. \ref{fig:topo_domain}-(c)). This point of view is supported by the observation that each of the topologically non-trivial domains hosts a pair of two gapless modes --- chiral edge modes --- with a negative and a positive group velocity at its left and right interfaces to the topologically trivial domains, respectively (see Fig. \ref{fig:boundary_positions}-(b)). Thus, the emergence of the observed local gapless modes dispersing parallel to the $y$-axis can be interpreted as a consequence of a bulk-edge correspondence although the modulation responsible for this behavior has a rather short length scale.\\
In this context, it is important that we have chosen a rather strong pairing interaction so as to have a coherence length shorter than the length scales of strain modulation. Only then a local picture as used here in our explanation in terms of topologically distinct domains has its validity, and thus the chiral subgap modes can be interpreted as edge modes of the chiral superconducting phases. This can be further supported by considering how the subgap modes shift for increased amplitude $ \delta t $ of the strain modulation, since the spatial location of the Lifshitz transition, i.e. position $x$ such that $\delta t_x = \delta t^*$, would move.
\begin{figure}[thpb]
	\vspace*{0.8cm}
	\hspace*{-0.7cm}
	\includegraphics[height=85mm]{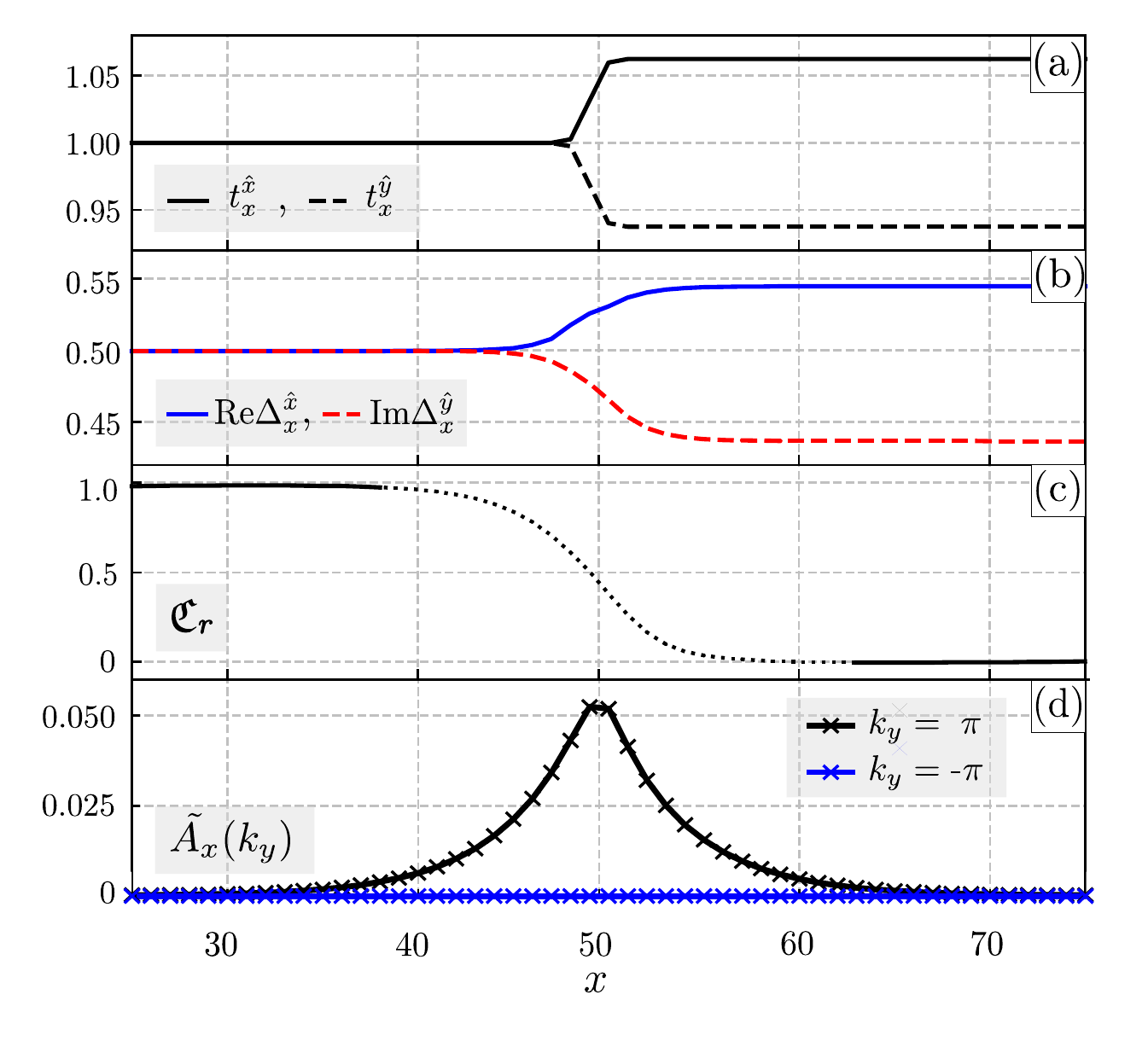}
	\vspace*{-0.2cm}
	\caption{Simulation of the modified model, where $N_x = 100, N_y = 41$ and $\delta t_x$ is defined in Eq. (\ref{eq:transfer_twoblock}). Presented are plots of (a) hopping amplitudes, (b) self-consistent order parameters $\Delta^{\hat x}_x$ and $\Delta^{\hat y}_x$, and (c) Local Chern markers (LCM), where non-integer values are observed (dotted curves) at the interface of the two domains. Note that mapping from Chern numbers to LCM generally fails near boundaries and interfaces of systems because of the way the map is constructed \cite{bianco2011mapping}. (d) Plot of the integrated spectral weight defined in Eq. (\ref{eq:integrated_spectral_fucntion}) showing a gapless state around $k_y = \pm\pi$ with positive velocity is localized around $x = 50$ where $\delta t_x = \delta t^*.$}
	\label{fig:twoblock_lcm}
\end{figure}
\paragraph{Local Chern Markers} --- To justify our discussion above, we will now apply the concept of the {\it local Chern markers} (LCMs), which essentially assigns to each position $(x,y)$ of a given system the Chern number that effectively reflects the local electronic structure at that position. The LCM was first introduced by Bianco and Resta for Chern insulators \cite{bianco2011mapping} in order to extend the notion of the Chern number to open and/or inhomogeneous systems, where the Chern number is in principle ill-defined due to the lack of translation invariance (an alternative approach to systems without translation symmetry was put forward in Ref. \cite{morimoto2015anderson}).  
As we will see below that the LCM has properties which can be applied well to topological superconducting systems like ours.\\
We define the LCM $\mathfrak{C}
_{\bm r}$ as a real-space dependent quantity by
\begin{equation}
\mathfrak{C}
_{\bm r} := -4\pi \text{Im} \langle \bm r | \hat{P}\hat{x}\hat{P}\hat{y}\hat{P}|\bm r \rangle, \label{eq:local_chern_marker_operators}
\end{equation}
where $\bm r = (x,y)\in \Lambda$. Here, $\hat{P}$ is the projection operator onto occupied Bogolyubov quasi-particle states and the set $\{|\bm r \rangle\}_{\bm r \in \Lambda}$ forms an orthonormal basis of single particle states in the Nambu representation that simultaneously diagonalize position operators $\hat{x}$ and $\hat{y}$. More detailed information of LCMs is given in the Appendix A. 

Now, we use LCMs to show that the modulated strain fields could induce topologically distinct domains in a single superconducting system like our model above. For this purpose, we focus on a region in the system where $\delta t$ crosses $\delta t^*$ and evaluate $\mathfrak{C}_{\bm r}$. Thus, we here employ a system of size $N_x = 100$ and $N_y = 41$ with Neumann periodic boundary conditions imposed along the $y$-direction. In addition, we introduce the following spatial variation of $\delta t_x$:
\begin{align}
\delta t_{x} =  \delta t^* \left\{ 1 + \tanh \left[ \frac{1}{2\ell} (2x- N_x) \right] \right \},
\label{eq:transfer_twoblock}
\end{align}
such that $ \delta t_x = \delta t^*$ at $x=N_x/2 $ ($ x = 0 $ corresponds to the undeformed lattice and $ x = N_x $ does to uniaxial compressed along $x$-direction) and the spatial variation of $ \delta t_x $ occurs on length scale of $\ell = 2 $ . By doing so, we implement at $x \sim N_x/2$ a boundary of two topologically distinct domains with $\mathcal{C} = 1 $ for $ x < N_x/2 $ and $ \mathcal{C}=0 $ for $ x > N_x/2$.
The numerical simulations of this system are given in Fig. \ref{fig:twoblock_lcm}, where the panel (a) shows the spatial variation of the hopping matrix elements and the panel (b) displays the self-consistent solution of the two order parameter components. The order parameter is shown to be chiral everywhere. However, the topology of the chiral superconducting state on the two sides of the boundary is different, as implied by the spatial change of $\mathfrak{C}_{\bm r}$ across $x = N_x/2$ as displayed in Fig. \ref{fig:twoblock_lcm}-(c). Our simulation also shows the presence of a chiral mode localized around $x = N_x/2$ with characteristic momentum $k_y = \pm \pi$ (see Fig. \ref{fig:twoblock_lcm}-(d)).
\subsection{CONCLUSIONS}
Our numerical study shows that uniaxial strain modulations on a chiral $p$-wave superconducting phase can lead to several intriguing features. One aspect is that the inhomogeneity of the order parameter yields spontaneous supercurrents in specific location. These supercurrents induced by these order parameter modulations likely provide better conditions for observation by scanning magnetic field probes than the ones at sample surfaces where surface roughness tends to suppress currents \cite{etter2018spontaneous}. A further important feature is the fact that modulated uniaxial strain sufficient to induce Lifshitz transitions between closed and open Fermi surfaces may also generate domains of different topology as indicated by the local Chern marker, which allows us to discriminate between domains of different topology. As a consequence the boundaries would carry gapless chiral quasiparticle modes. In the example discussed here, the Chern number changes between 1  to 0 yielding exactly one gapless chiral quasiparticle mode at the boundary. This is different from the situation we encounter at chiral domain walls between phases of opposite chirality with Chern numbers $+1$ and $-1$, which then host two chiral modes. The presence of such modes would possibly be detectable by local spectral probes such as STM/STS. The control on spatial strain modulation could also be a way to induce spatial topological phase transitions in single-crystal samples. In this way the fabrication of samples with strain modulations may be a fruitful platform to probe unconventional superconducting order parameters. 

\begin{acknowledgments}
	\section{Acknowledgements}
	The authors thank to Philip Moll, Mark Fischer, Aline Ramires, Christopher Mudry, and Martina Soldini for many helpful discussions. 
	This work was financed through the Swiss National Science Foundation (Ambizione Grant No. 186043 and Division II Grant No. 184739).
\end{acknowledgments}

\appendix
\section{APPENDIX A: LOCAL CHERN MARKERS}
Chern numbers are known to be ill-defined when systems lack translation invariance. As an extension of the notion of the Chern numbers to open and/or inhomogeneous systems of two-dimensional Chern insulators, Bianco and Resta introduced the notion of local Chern markers (LCMs) \cite{bianco2011mapping}. Local Chern marker operator $\hat{\mathfrak{C}}$ is defined by
\begin{equation}
\hat{\mathfrak{C}} := -4\pi \text{Im} \left( \hat{P}\hat{x}\hat{P}\hat{y}\hat{P} \right), \label{eq:local_chern_marker_operators_appendix}
\end{equation}
where $\hat{P}$ denotes the projection operator onto occupied states of the elementary fermionic excitation of a given system and $\hat{x}$ and $\hat{y}$ are the usual position operators. In the following, we consider systems of spinless electrons for simplicity.

In the case of infinite systems with translation invariance and periodic boundary conditions, the Chern number $\mathcal{C}$ and $\hat{\mathfrak{C}}$ are related as follows:
\begin{align}
	\mathcal{C} :=&\; -\frac{1}{\pi}\text{Im} \sum_{m} \int_\text{1st-BZ} \text{d}k_x \text{d}k_y \langle \Psi_{m\bm k}| \hat{P}\hat{x}\hat{P}\hat{y} |\Psi_{m\bm k}\rangle\\
	=&\; -\frac{1}{\pi}\frac{(2\pi)^2}{A_\text{cell}}\text{Im} \text{Tr}_{\Lambda_\text{cell}}\left( \hat{P}\hat{x}\hat{P}\hat{y}\hat{P} \right)\\
	=&\;	 \frac{1}{A_\text{cell}}\text{Tr}_{\Lambda_\text{cell}}\left(\hat{\mathfrak{C}}\right), \label{eq:Chern_trace}
\end{align}
where $A_\text{cell}$ is the area of a unit cell $\Lambda_\text{cell}$ of the lattice in question. The sum in the first line is taken over quantum numbers $m$ of Bloch functions. Note that the trace is invariant under basis transformation from the Bloch basis $\{ |\Psi_{m\bm k}\rangle\}$ to the position basis $\{ |\bm r \rangle\}_{\bm r \in \Lambda_\text{cell}}$. It thus follows from Eq. (\ref{eq:Chern_trace}) that
\begin{align}
	\mathcal{C}
	= \frac{1}{A_\text{cell}}\text{Tr}_{\Lambda_\text{cell}}(\hat{\mathfrak{C}})
	&= \frac{1}{A_\text{cell}}\sum_{\bm r \in \Lambda_\text{cell}} \langle \bm r| \hat{\mathfrak{C}} | \bm r \rangle\\
	&= \frac{1}{A_\text{cell}}\sum_{\bm r \in \Lambda_\text{cell}} \mathfrak{C}_{\bm r},
	\label{eq:chern_and_LCM}
\end{align}
where we define the local Chern marker (LCM) $\mathfrak{C}_{\bm r}$ as
\begin{equation}
\mathfrak{C}_{\bm r} := -4\pi \text{Im} \langle \bm r | \hat{P}\hat{x}\hat{P}\hat{y}\hat{P} | \bm r \rangle.
\label{eq:LCM_def}
\end{equation}
Note that $\mathcal{C}$ in Eq. \eqref{eq:chern_and_LCM} does not depend on the particular unit cell in real space that is chosen to take the trace. Thus, with a supercell $\Lambda \supset \Lambda_{\text{cell}}$ of area $A\, (> A_{\text{cell}})$, the following equality holds:
\begin{equation}
    \frac{1}{A_\text{cell}}\sum_{\bm r \in \Lambda_\text{cell}} \mathfrak{C}_{\bm r} = \frac{1}{A}\sum_{\bm r \in \Lambda} \mathfrak{C}_{\bm r} =: \langle \mathfrak{C} \rangle_{\Lambda}, \label{eq:lcm_superlattice}
\end{equation}
where $\langle \mathfrak{C} \rangle_{S}$ is the average of LCMs over a subsystem $S$. Since the supercell $\Lambda$ itself is finite, it is known to hold that $\langle \mathfrak{C} \rangle_{\Lambda} = 0$. Thus, $\langle \mathfrak{C} \rangle_{\Lambda}$ is not a good indicator of the topology of local electronic structure. However, one may split the supercell into its edge $\partial \Lambda$ and its bulk $\interior{\Lambda}$ parts and accordingly express $\langle \mathfrak{C} \rangle_{\Lambda}$ as
\begin{equation}
    \langle \mathfrak{C} \rangle_{\Lambda} = \langle \mathfrak{C} \rangle_{\partial \Lambda} + \langle \mathfrak{C} \rangle_{\interior{\Lambda}}.
\end{equation}
From previous numerical studies on Chern insulators, it has been suggested that \cite{bianco2011mapping, caio2019topological, irsigler2019microscopic}
\begin{equation}
    \lim_{A \rightarrow +\infty} \langle \mathfrak{C} \rangle_{\interior{\Lambda}} = \mathcal{C}.
\end{equation}
Note that the definition of the average of LCMs over a subsystem, which is in general open and inhomogeneous within, relies on neither periodic boundary conditions nor translation invariance of the infinite system to have started our discussion with. Thus, Eq. \eqref{eq:lcm_superlattice} is applicable to finite systems and/or a subsystem of an infinite systems so as to evaluate the topology of its (local) electronic structure.\\
Lastly, with the resolution of identity $\sum_{\bm r \in \Lambda} |\bm r \rangle \langle \bm r| = \hat{1}$ and single particle density matrix $\rho_{(\bm r, \bm{r'})}$ of spinless electrons, Eq. (\ref{eq:LCM_def}) can be expressed as
\begin{equation}
\mathfrak{C}_{\bm r} = -4\pi \text{Im} \sum_{\bm {r'}, \bm {r''} \in \Lambda} \rho_{(\bm r, \bm{r'})} x' \rho_{(\bm{r'}, \bm{r''})} y'' \rho_{(\bm{r''}, \bm r)}, \label{eq:local_chern_marker_density_matrix}
\end{equation}
with
\begin{align}
\rho_{(\bm r, \bm {r'})} &:= \sum_{n\,:\, \epsilon_n < \epsilon_{\text{F}}} \psi_{n,\bm r} \psi^*_{n,\bm{r'}} = \langle \bm r | \hat{P} | \bm r' \rangle,
\end{align}
and
\begin{align}
|\bm r \rangle &:= \hat{c}^\dagger_{\bm r} | 0 \rangle,\\
| n \rangle &:= \hat{b}^\dagger_{n} | 0 \rangle,\\
\psi_n(\bm r) &:= \langle \bm r | n \rangle,
\end{align}
where $\hat{c}^\dagger_{\bm r}$ ($\hat{c}_{\bm r}$) is the creation (annihilation) operator of a spinless electron at site $\bm{r} = (x,y) \in \Lambda$, and $\hat{b}^\dagger_{n}$ ($\hat{b}_{n}$) is the creation (annihilation) operator of an electron in eigenstate $n$ of a given total Hamiltonian with eigenenergy $\epsilon_n$. $\epsilon_F$ stands for the Fermi energy, and $|0\rangle$ is the vacuum state, which satisfies $\hat{c}^\dagger_{\bm r} \hat{c}_{\bm r}|0 \rangle = \hat{b}^\dagger_{n} \hat{b}_{n}|0 \rangle = 0$ and $\langle 0 | 0 \rangle = 1$.\\

We now aim at extending the notion of LCMs to systems with BdG Hamiltonians in the Nambu representation. We observe that the Nambu representation doubles the dimension of the original Hilbert space in question. Thus, the position basis over which the trace in Eq. (\ref{eq:chern_and_LCM}) is taken also needs to double its dimension by taking the union of two position bases with respect to particles (P) and holes (H). Hence, $\{ |\bm r \rangle \}_{\bm r \in \Lambda}$ should read
\begin{align}
	\{ |\bm r \rangle \}_{\bm r \in \Lambda} = \{ |\bm r_{P} \rangle \}_{\bm r \in \Lambda} \cup \{ |\bm r_{H} \rangle \}_{\bm r \in \Lambda},
\end{align}
with
\begin{align}
|\bm r_P \rangle &:= \hat{c}^{\dagger}_{\bm r} | 0 \rangle \neq 0,\\
|\bm r_H \rangle &:= \hat{c}_{\bm r} |\bar{0} \rangle \neq 0,
\end{align}
where $|\bar{0}\rangle$ is the ground state with respect to holes, i.e. the fully occupied state. Note that $\langle 0| \bar{0} \rangle = 0.$ Thus, we redefine LCMs for superconducting systems as
\begin{equation}
\mathfrak{C}_{\bm r} := \sum_{I \in \{ {\text P, H} \} } \mathfrak{C}^{I}_{\bm r} \label{eq:local_chern_marker_sc},
\end{equation}
with
\begin{equation}
\mathfrak{C}^{I}_{\bm r} := -4\pi \text{Im} \sum_{I', I'' \in \{ \text{P}, \text{H}\}} \sum_{\bm {r'}, \bm {r''} \in \Lambda} \rho^{I, I'}_{(\bm r, \bm{r'})} x' \rho^{I',I''}_{(\bm{r'}, \bm{r''})} y'' \rho^{I'',I}_{(\bm{r''}, \bm r)},
\end{equation}
and
\begin{align}
\rho^{I,I'}_{(\bm r, \bm {r'})} :=& \sum_{n\,:\,\epsilon_n < \epsilon_{\text{F}}} \psi^{I'}_{n, {\bm r'}} \psi^{I\;*}_{n,\bm r} = \sum_{n\,:\,\epsilon_n < \epsilon_{\text{F}}}  \langle \bm r'_{I'} | n \rangle \langle n | \bm r_I \rangle, \label{eq:single_particle_density_matrix}
\end{align}
where $n$ labels eigenstates of Bogolyubov quasi-particles and $|n\rangle$ is the $n$-th eigenstate with eigenenergy $\epsilon_n$.
\section{APPENDIX B: BOGOLYUBOV TRANSFORMATIONS}
Here we show the derivation of the expression of the single particle density matrices in Eq. \eqref{eq:single_particle_density_matrix} in terms of coherence factors in the Bogolyubov transformations.\\

We consider a BdG mean field Hamiltonian $\hat{\mathcal{H}}_{\text{MF}}$ in the Nambu representation that describes electrons on an $N_x \times N_y$ square lattice $\Lambda := \{ (x, y) \in \mathbb{N} \times \mathbb{N}\, | \, 0 \leq x < N_x, \, 0 \leq y < N_y \}$ and is written in the following bilinear form:
\begin{align}
\hat{\mathcal{H}}_{\text{MF}}
&= \hat{\bm{\Psi}}^\dagger H_{\text{MF}} \hat{\bm{\Psi}},
\end{align}
with Nambu spinor $\hat{\bm \Psi}$,
\begin{equation}
\hat{\bm \Psi} := \begin{pmatrix}
\hat{c}_{\underline{0}}\\
\hat{c}^\dagger_{\underline{0}}\\
\hat{c}_{\underline{1}}\\
\hat{c}^\dagger_{\underline{1}}\\
\vdots\\
\hat{c}_{\underline{x+y\cdot N_x}}\\
\hat{c}^\dagger_{\underline{x+y\cdot N_x}}\\
\vdots\\
\hat{c}_{\underline{N_\text{tot}-1}}\\
\hat{c}^\dagger_{\underline{N_\text{tot}-1}}\\
\end{pmatrix}
=
\begin{pmatrix}
\hat{c}_{0,0}\\
\hat{c}_{0,0}^{\dagger}\\
\hat{c}_{1,0}\\
\hat{c}_{1,0}^{\dagger}\\
\vdots\\
\hat{c}_{x,y}\\
\hat{c}_{x,y}^{\dagger}\\
\vdots\\
\hat{c}_{N_x-1,N_y-1}\\
\hat{c}_{N_x-1,N_y-1}^{\dagger}\\
\end{pmatrix},
\end{equation}
where $\hat{c}^\dagger_{\underline{r}}$ and $\hat{c}_{\underline{r}} \, (\underline{r}\in \mathbb{N} : 0 \leq \underline{r} < N_\text{tot})$ are the creation and annihilation operators of a spinless electron at site $\bm r = (\underline{r} \mod N_x, \,\text{floor}(\underline{r}/N_x)) \in \Lambda$, respectively, and $N_{\text{tot}} := N_x\times N_y$. $H_{\text{MF}}$ is a Hermitian matrix of dimension $2N_{\text{tot}}$ that is in general not yet diagonalized. Being Hermitian, $H_{\text{MF}}$ can be diagonalized by the similarity transformation with a unitary matrix $U$ of dimension $2N_{\text{tot}}$. The components of $U$ are called coherence factors and hereafter denoted by
\begin{align}
(U^\dagger)_{n,\underline{2r}} &= v_{n,\underline{r}} \;\;\;\;\; \text{(hole-like)},\\
(U^\dagger)_{n,\underline{2r+1}} &= u_{n,\underline{r}} \;\;\;\;\; \text{(particle-like)},
\end{align}
where $n \in \{ m \in \mathbb{N} \,|\, 0 \leq m < N_{\text{tot}} \}$. Then, it follows
\begin{align}
\hat{\mathcal{H}}_{\text{MF}}
&= \hat{\bm{\Psi}}^\dagger H_{\text{MF}} \hat{\bm{\Psi}}\\
&= (\hat{\bm{\Psi}}^\dagger U)
\underbrace{(U^\dagger H_{\text{MF}} U )}_{=:H_\text{B}}
\underbrace{(U^\dagger \hat{\bm{\Psi}})}_{=:\hat{\bm{\Psi}}_\text{B}}\\
&= (\hat{\bm{\Psi}}_\text{B})^\dagger H_\text{B} \hat{\bm{\Psi}}_\text{B},
\end{align}
where $\hat{\bm \Psi}_\text{B}$ is composed of the creation (annihilation) operators of Bogolyubov quasi-particles $\hat{b}^\dagger_n$ ($\hat{b}_n$) in the $n$-th eigenstate of $\hat{H}_{\text{MF}}$ with eigenenergy $E_n$:
\begin{align}
\hat{\bm\Psi}_\text{B}
:=&\; U^\dagger \hat{\bm{\Psi}}\\
 =&\; (\hat{b}_0, ... , \hat{b}_n, ... , \hat{b}_{N_\text{tot}-1}, \hat{b}^\dagger_{N_\text{tot}-1}, ..., \hat{b}^\dagger_n, ..., \hat{b}^\dagger_0)^\text{T}, \label{App:Eq:nambu_bogo_basis}
\end{align}
and a diagonal matrix $H_{\text{B}}$ whose matrix elements accordingly correspond to the eigenenergies of Bogolyubov quasi-particles:
\begin{align}
H_{\text{B}} = \text{diag}(-E_0, ... , -E_n, ..., E_n, ... , E_0),
\end{align}
with $E_{m} \geq 0$ for all $m$. The transformation Eq. \eqref{App:Eq:nambu_bogo_basis} defines the Bogolyubov transformation between quasi-particle operators $\hat{b}^\dagger_n, \hat{b}_n$ and electron operators $\hat{c}^\dagger_{\underline{r}}, \hat{c}_{\underline{r}}$, whose explicit expression is
\begin{align}
\hat{b}_n
=&\; \sum^{N_\text{tot}-1}_{\underline{r}=0} \left\{ 
(U^\dagger)_{n,\underline{2r}} \hat{c}
^\dagger_{\underline{r}} + (U^\dagger)_{n,\underline{2r+1}} \hat{c}_{\underline{r}} \right\}\\
=&\; \sum^{N_\text{tot}-1}_{\underline{r}=0} \left(v_{n,\underline{r}} \hat{c}^\dagger_{\underline{r}} + u_{n,\underline{r}} \hat{c}_{\underline{r}} \right),\\
\hat{b}^\dagger_n
=&\; \sum^{N_\text{tot}-1}_{\underline{r}=0} \left( v^*_{n,\underline{r}} \hat{c}_{\underline{r}} + u^*_{n,\underline{r}} \hat{c}^\dagger_{\underline{r}} \right),
\end{align}	
and whose inverse transformation is given by
\begin{align}
	\hat{c}_{\underline{r}} &= \sum^{N_\text{tot}-1}_{n=0}
	\left(
	v^*_{n,\underline{r}}\hat{b}_n + u_{n,\underline{r}}\hat{b}^\dagger_n
	\right),\\
	\hat{c}^\dagger_{\underline{r}} &= \sum^{N_\text{tot}-1}_{n=0}
\left(
u^*_{n,\underline{r}}\hat{b}_n + v_{n,\underline{r}}\hat{b}^\dagger_n
\right),
\end{align}
where we use the relations $v_{N-n,\underline{r}} = u^*_{n,\underline{r}}$ and $u_{N-n,\underline{r}} = v^*_{n,\underline{r}}$ where $N := 2N_\text{tot} - 1$, which are imposed by the fact that $U$ is unitary.\\

\paragraph{Single Particle Density Matrices} --- The single particle density matrices $\rho^{I,I'}_{(\bm r, \bm {r'})}$ in Eq. (\ref{eq:single_particle_density_matrix}) are expressed as follows:
\begin{align}
\rho^{P,P}_{(\bm{r}, \bm{r'})}
&=\sum^{N_\text{tot}-1}_{n=0} u^*_{n,\underline{r}} u_{n,\underline{r'}},\\
\rho^{P,H}_{(\bm{r}, \bm{r'})}
&=\sum^{N_\text{tot}-1}_{n=0} u^*_{n,\underline{r}} v_{n,\underline{r'}} = \left(\rho^{H,P}_{(\bm{r}, \bm{r'})}\right)^*,\\
\rho^{H,H}_{(\bm r, \bm {r'})}
&= \sum^{N_\text{tot}-1}_{n=0} v^*_{n,\underline{r}} v_{n,\underline{r'}},
\end{align}
where $\underline{r}$ satisfies $\bm r = (\underline{r} \mod N_x, \,\text{floor}(\underline{r}/N_x))$.\\

\bibliography{reference}

\end{document}